\documentclass[iop,usenatbib]{emulateapj}

\usepackage{graphicx}
\usepackage{journals}
\usepackage{amsmath}
\usepackage{color}

\shorttitle{IMF from the network theory}
\shortauthors{Klishin \& Chilingarian}

\begin{document}

\title{Explaining the stellar initial mass function with
the theory of spatial networks}

\author{Andrei A.~Klishin\altaffilmark{1,2,*}}
\author{Igor Chilingarian\altaffilmark{3,4,*}}
\altaffiltext{1}{Department of Physics, Massachusetts Institute of Technology, 77 Massachusetts Ave, Cambridge MA 02139, USA}
\altaffiltext{2}{Department of Physics, University of Michigan, Ann Arbor,
Michigan 48109, USA}
\altaffiltext{3}{Smithsonian Astrophysical Observatory, 60 Garden St. MS09, Cambridge MA 02138, USA}
\altaffiltext{4}{Sternberg Astronomical Institute, Moscow State University, 13 Universitetsky prospect, Moscow, 119992, Russia}
\email{$^*$ e-mail: aklishin@umich.edu, igor.chilingarian@cfa.harvard.edu}

\date{\today}


\begin{abstract}
The distributions of stars and prestellar cores by mass (initial and dense
core mass functions, IMF/DCMF) are among the key factors regulating star
formation and are the subject of detailed theoretical and observational 
studies. 
Results from numerical simulations of star formation qualitatively resemble
an observed mass function, a scale-free power law with a sharp decline at low
masses.  However, most analytic IMF theories critically depend on the
empirically chosen input spectrum of mass fluctuations which evolve into
dense cores and, subsequently, stars, and on the scaling relation between
the amplitude and mass of a fluctuation.  Here we propose a new approach
exploiting the techniques from the field of network science. We represent a system of dense
cores accreting gas from the surrounding diffuse interstellar medium (ISM)
as a spatial network growing by preferential attachment and assume that the
ISM density has a self-similar fractal distribution following the Kolmogorov
turbulence theory. We effectively combine gravoturbulent
and competitive accretion approaches and predict the accretion rate to be
proportional to the dense core mass: $dM/dt \propto M$. Then we describe 
the dense core growth and demonstrate that the power-law core mass function 
emerges independently of the initial distribution of
density fluctuations by mass.  Our model yields a power law solely defined by the fractal
dimensionalities of the ISM and accreting gas. With a proper choice of the
low-mass cut-off, it reproduces observations over three decades in
mass. We also rule out a low-mass star dominated ``bottom-heavy'' IMF in a
single star-forming region.
\end{abstract}


\section{Introduction and Motivation}

Six decades ago the stellar initial mass function (IMF) was derived from star 
counts
\citep{Salpeter55} as a scale-free power law ($dN/dm \approx m^{\alpha}$;
$\alpha = -2.35$) with more frequent low-mass stars than high-mass stars. 
Since then, it has attracted attention as one of the principal star formation
characteristics that controls stellar feedback and, therefore, governs
galaxy evolution.  Furthermore, explaining the IMF will help us to
understand the star formation physics.  From observations, the IMF shape
appears to be universal across different star-forming regions
\citep{Kroupa02}.  Resembling a unimodal \citep{Salpeter55} or bimodal
\citep{Kroupa01} power law or a log-normal distribution with a power-law
tail \citep{Chabrier03}, it sharply declines at the low end at masses
$<1/12$ solar mass ($M_{\odot}$).  The dense core mass function (DCMF) derived 
from observations of
giant molecular clouds \citep{ALL07,Andre+10} is an IMF precursor: First,
dense cores grow from density fluctuations by attracting surrounding
material, cool down, then protostars form inside them and evolve into stars. 
The DCMF shape also looks like a power law at high masses and declines below
1/3~$M_{\odot}$, offset by a factor of $\sim 4$ to higher masses compared to
the stellar IMF, illustrating a $\sim$25\%\ gas-to-stars transformation
efficiency.  Thus, if robust arguments were provided to explain the DCMF
shape, the IMF shape would follow through that heuristic conversion rule.

All existing analytic and numerical IMF theories (see the review by
\citealp{HC11}) consider either the accretion of material on protostars
\citep{Zinnecker82,BB06} or the gravitational fragmentation of the 
interstellar medium (ISM)
\citep{PN02}.  In a simple model, a nonlinear stage of the molecular cloud
fragmentation yields a low-mass IMF decline \citep{ST79} but does not
reproduce a power-law high-mass tail.  Diverse physical mechanisms of the
molecular cloud cooling that affect the collapse and fragmentation are often
hidden in a complex equation of state for molecular clouds where the
polytropic index depends on density, temperature, and chemical composition
\citep{SS00}. One of the most complete analytic IMF theories up to date
\citep{HC08,HC13} explains the observed overall IMF shape by analyzing the
evolution of density fluctuations in a self-gravitating turbulent ISM in the
presence of a magnetic field.  Similarly to other gravoturbulent theories,
in order to reproduce a power-law part of the mass function consistent with the
Salpeter slope, it relies on a specific choice of the scaling relation
that connects the density fluctuation amplitude with the mass contained within
that fluctuation, and a log-normal initial
ISM density probability density function (PDF) required for the analytic computation of the
mass function shape.
However, the density PDF scale dependence chosen in \citet{HC08} relies
on results from numerical simulations. Also, density PDFs observed in
molecular clouds deviate substantially from the log-normal shape and 
vary across different star-forming regions \citep{LAL15}.  Moreover,
gravoturbulent theories do not consider any external accretion on dense
core progenitors, so there is no guarantee that a power-law DCMF holds as
the system evolves.

Recent developments by \citet{Hopkins2013} introduce a variety of
modifications to the gravoturbulent fragmentation conditions and propose
different grounds for the density PDF scaling relation.  The author
claims that the input ISM density distribution does not have to be
log-normal.  However, this might be the result of the mathematical
simplification applied (the top-hat filtering in the Fourier space and the
consequent Taylor expansion) that transforms an arbitrary function into a
log-normal like shape.  Nevertheless, this theory yields a fluctuation
mass function consistent with the Salpeter slope after some fine tuning
of the model parameters.  At
the same time, no argument is provided for the fundamental reasons of the
power law and a particular value of the exponent.

The generalization of gravoturbulent theories to non-log-normal or
non-analytic PDF shapes becomes cumbersome and not very straightforward. 
\citet{SKFK10} demonstrate analytically and numerically that
gravoturbulent theories can reproduce the power-law IMF for arbitrary
initial PDF shapes, however, in their analytic computation they rely on
specific simplifications (e.g.  excluding the largest cores from
consideration in the case of the Hennebelle--Chabrier theory), which
\emph{de facto} restricts the input density PDFs to certain functional
families.

Whereas analytic theories that deal with the competitive accretion scenario
reproduce the power-law IMF tail, they fail to match its observed exponent
$(\alpha=-2.35)$ and cannot be applied to non-clustered star formation. 
A recent IMF theory from that family \citep{Basu2015} generates
a log-normal distribution with a power-law tail via the
process of quenched accretion with the exponential distribution of 
accretion timescales.  However, that theory neither provides a justification
for the exponential distribution nor does it proposes a quantitative argument
for the slope of the power-law tail. 

The theory by \cite{Maschberger13} considers both linear and nonlinear
accretion of mass onto dense cores and takes into account the stochasticity
of the process using the Stratonovich stochastic calculus.  While it
successfully reproduces high-mass power-law tails, the accretion laws
considered there are not connected to the ISM properties.

Hence, the following questions still remain unanswered by existing IMF
theories: (i) an a scale-free power-law distribution of dense core masses
become established by some physical processes independent of the initial
density distribution; (ii) does it hold as the system evolves by accreting
external material; and (iii) what does the power-law exponent depend on?

Here we present a different approach to the analytic DCMF theory.  We describe the
mass accretion onto prestellar dense cores in the fractal ISM as
preferential attachment, a key phenomenon studied in the field of \emph{network
science} \citep{Merton68,newman2010networks,Barthelemy11}.  We use
probabilistic accounting of small parcels that join dense cores, subject to
gravitational attraction and stochastic noise.  We limit our model to 
the early stages of the dense core growth by mass accretion without complex
physics following the protostar formation.  Therefore, it applies to
starless dense cores.  The principal result of our theory is an analytic
expression for the power-law tail that develops for \emph{any} initial 
distribution of dense core
progenitors by mass. This is the first example of a theory that
consistently connects the accretion rate to global properties of the
turbulent ISM and, thus, effectively combines the gravoturbulent and
competitive accretion approaches.

\begin{figure}
\hskip -20pt
\includegraphics[width=0.6\hsize]{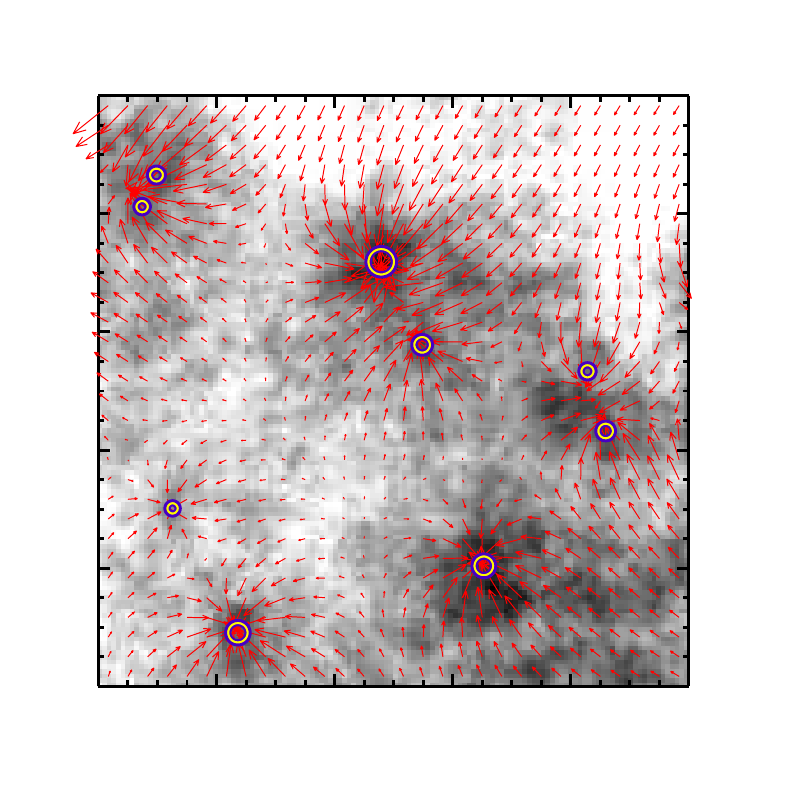}
\hskip -32pt
\includegraphics[width=0.6\hsize]{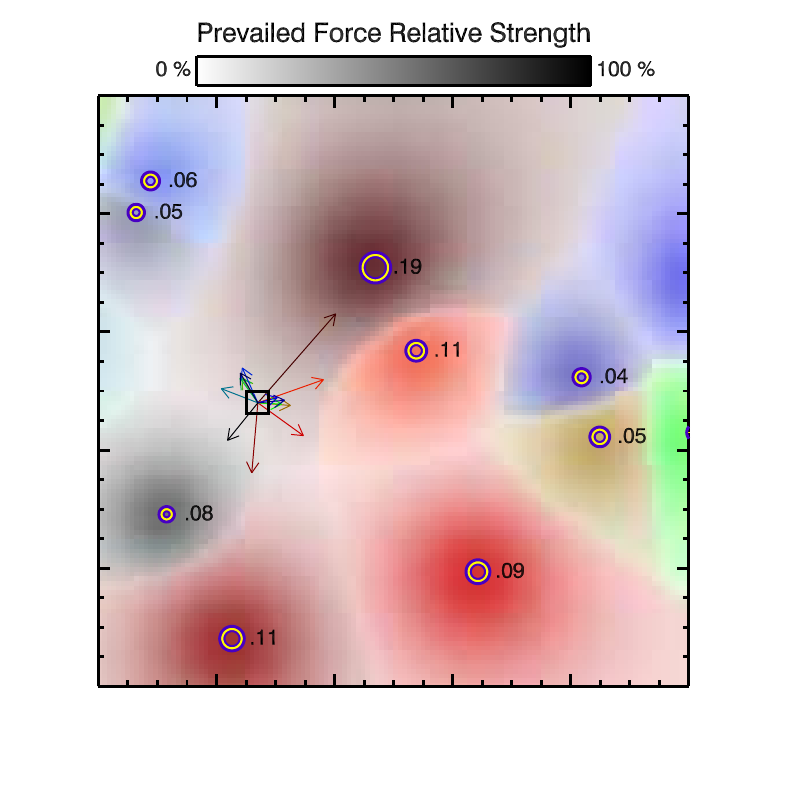}
\caption{({\bf Left}) We generated a 3-dimensional fractal density field
\citep{Elmegreen97} with the dimensionality $D_m=2.35$, projected it onto a
plane, and identified dense core progenitors as overdensities shown as
circles with radii proportional to masses.  A
gravitational acceleration field generated by dense cores is displayed by
vectors.  ({\bf Right}) In the same system of dense cores,
the color density corresponds to the the fraction (0 to 1) of the
prevailing force in the overall force balance, while different colors stand
for different basins of attraction.  The lengths of vectors from the parcel
(box) show the accretion probabilities by corresponding dense cores.
\label{fig_model}}
\end{figure}

\begin{figure}
\includegraphics[width=\hsize]{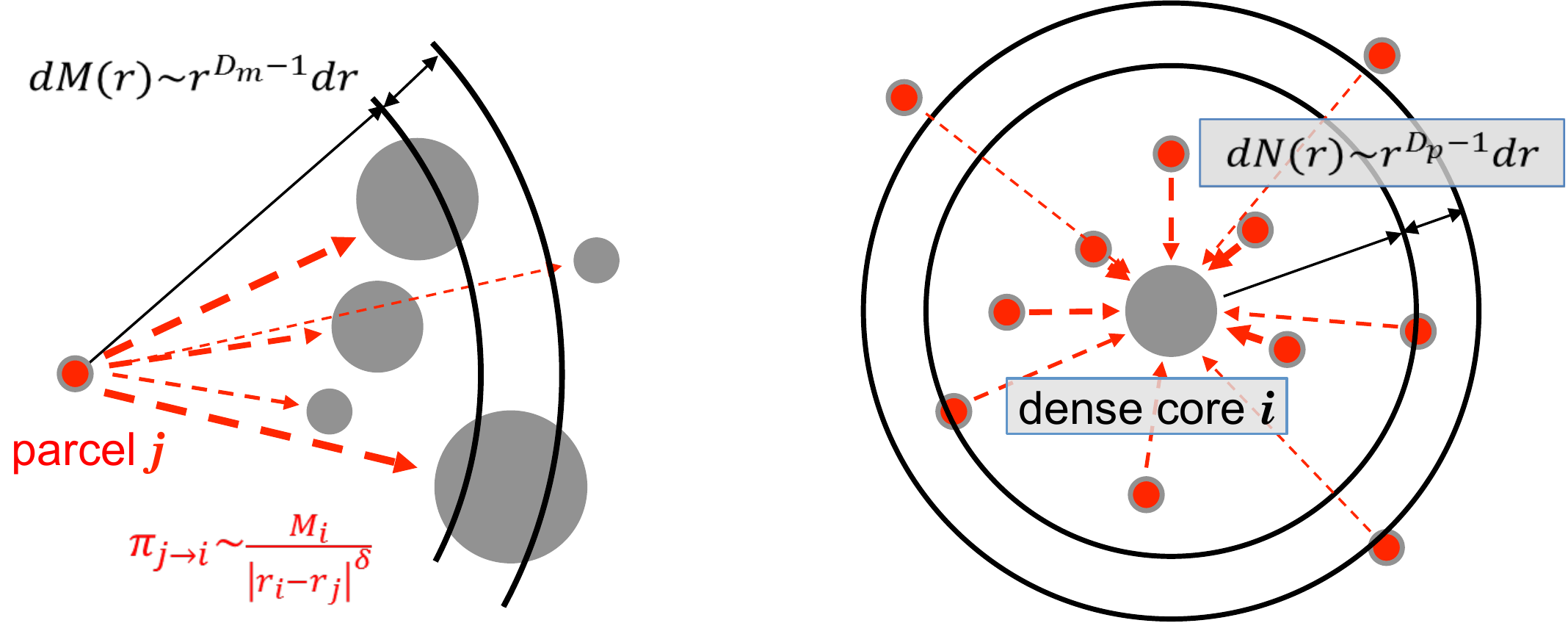}
\caption{({\bf Left}) A new parcel $j$ emerges in
an arbitrary point of the system and chooses between different fractally 
distributed dense cores $i$ it can attach to, with the attachment probability 
directly proportional to the gravitational acceleration. ({\bf Right}) The 
same process as seen by a core $i$: as new identical fractally distributed parcels 
emerge throughout the system, they can be accreted with probabilities
decaying with distance.
\label{fig_cartoon}}
\end{figure}

\section{Mass distribution and a system of units} 

We describe a system of dense cores as an array of masses $M_i$ distributed
according to some time-dependent function $p(m,t)$.  At a given moment, the
system has a total mass $M_\text{tot}$ and a total number of dense cores
$N$.  The mass distribution shape is governed by the two processes:
accretion of mass parcels onto existing cores and generation of new core
progenitors.  We first treat events of parcel accretion and, thus, masses of
dense cores as discrete and further take the continuum limit.

Let us consider an event where a gas parcel of a small mass $dm$ per one unit
of time $dt$ attaches to one of the cores.  Regardless of which core it
joins, the total mass of the system grows by $dm$, hence $dM_\text{tot}=dm$. 
The accretion summed over the entire system (e.g. the global mass
growth of the system) happens at some characteristic rate $K=dm/dt$.  Over a
time step $dt$ every mass bin $m$ can only be affected by bins within
$dm=Kdt$ from it.  Hence, the timescale choice uniquely defines the mass
bin size.  A volume density of new dense core progenitors created per unit
time in each bin is described by some function $f(m)$.  This function is the
only mechanism that increases $N$, therefore
$dN/dt=\int\limits_{0}^{\infty}f(m)dm=F$.  We stress that the $K$ and
$F$ values are, in general, time-dependent.  As the accretion exhausts
available material in the surrounding ISM, both rates should slowly decay to
zero.  However, as we show in Section 5, neither their exact time dependence
nor their absolute values matter for the final DCMF shape.

Then, the mass distribution of cores
is expressed as $p(m,t)$ with $m\in \{dm,2dm,3dm,...\}$, discrete time
$t\in \{dt,2dt,3dt,...\}$, and normalization
$\sum\limits_{m=1}^{\infty}p(m,t)=1$.  We are interested in its long term
behavior, such that $p(m,t)\underset{t\to
\infty}{\to}p(m)$.

\section{The fractal matter distribution in the ISM} 

Observations
suggest that density and velocity distributions in the ISM are predominantly
defined by turbulent motions on scales from hundredths of a parsec to
hundreds of parsecs \citep{ES04}.  A consequence of the Kolmogorov
theory \citep{Kolmogorov41} is a self-similar or fractal density distribution
in a turbulent flow \citep{SRM89} with the predicted fractal dimensionality
$D=7/3=2.33$.  It stays in agreement with measurements 
obtained from observations of giant molecular clouds
\citep{FPW91,EF96} ($D=2.2 - 2.4$) and in laboratory studies of turbulence
\citep{SRM89} ($D=2.35$).  Numerical simulations of the ISM evolution with
an input fractal density field \citep{Elmegreen97,Elmegreen02} yield
power-law mass distributions of overdensities that
correspond to dense cores in star-forming regions.

We describe ISM as a two-phase medium where the two phases may have
different fractal dimensions.  One phase corresponds to dense cores.  The
other one corresponds to parcels, small gas/dust fragments with individual
masses substantially below the turbulent and gravitational Jeans masses
\citep{HC08,HF12} that do not have to obey the same law.  Dense cores arise
from the initial turbulent medium and, therefore, their positions trace the
initial overdensities in the turbulent flow.  Parcels correspond to all
remaining material of the ISM that did not enter dense core overdensities
initially, but can be accreted by them.  The dense core phase defines the
gravitational field profile in the ISM, and the parcel phase moves in that
field and accretes on dense cores.  The spatial distributions of the two
phases are governed by different mechanisms and thus are neither positively
nor negatively correlated.

Our mathematical description of the two-phase ISM follows a fractal 
model by \cite{Tarasov05}. Both phases have a characteristic ``microscopic'' 
lengthscale (\textit{pore size} in \cite{Tarasov05}) on which the 
discreteness of the medium is visible. For the dense core phase this 
characteristic lengthscale $l$ corresponds to the average distance between  
adjacent dense cores. It can be inferred from the Jeans mass and the total 
system mass. For the parcel phase, the lengthscale $d$ corresponds to the 
average separation between particles of gas or dust. Presumably $l\gg d$, 
which corresponds to dense cores being much larger and less frequent than 
parcels. However, for the purpose of this calculation we are interested in 
the statistics on much larger scales.

As \cite{Tarasov05} suggests, for a fractal medium observed on scales
$r\gg l$ or $r\gg d$, uniform fractal scaling is observed, i.e.  the mass
confined in a sphere of radius $r$ grows as $M(r)\sim r^{D_m}$, where $D_m$
is the fractal dimensionality of the density distribution.  This relation
has two important properties, \textit{fractality} and \textit{homogeneity}. 
For the \textit{uniform} distribution of matter in 3D space $D_m=3$, but for
the \textit{fractal} distribution $D_m<3$.  Consequently, the average
density in a sphere $\rho \sim M(r)/r^3$ is not constant, but is
scale-dependent.  The \textit{homogeneity} means that the fractal power-law
scaling of mass confined in a sphere is independent of the position of that
sphere.  Features and structures of the fractal distribution are associated
with observing it on different scales rather than at different positions. 
Self-similar (or fractal) scaling appears only on scales when discrete features
of characteristic scales $l$ and $d$ blur out.

Since the dense core distribution traces the initial Kolmogorov-like 
supersonic turbulent distribution, we take it to have $D_m=7/3\simeq 2.33$. 
Confining our
system to a box of linear size $L$ and normalizing its mass to
$M_\text{tot}$, the mass confined in a thin concentric sphere becomes
$dM(r)=M_\text{tot} D_m r^{D_m-1} L^{-D_m} dr$. The fractal dimension of 
parcels $2<D_p\leq 3$ is left as a free parameter for now and is discussed 
below in more detail. Therefore, the average number of
parcels confined within a thin concentric sphere, properly normalized, is
$dn(r)=D_p r^{D_p-1} L^{-D_p} dr$.

Even if a substantial mass fraction is contained in the diffuse phase
(e.g.  an order of 50\%), our calculations will remain valid, because the
gravitational field gradients and, correspondingly, the basins of attraction
(see Fig.~\ref{fig_model} right) will still be defined by ``point-like''
dense cores.

\section{Dense core growth by preferential attachment} 

Preferential
attachment is a stochastic process in which a set of objects possessing some
property acquire discrete units of this property in a partly random fashion
such that the probability of a given unit to be attached to a given object
increases with the increase of the amount of that property already contained
in this object.  It is also referred as a ``Yule process'' in speciation \citep{Yule25}, a
``Matthew effect'' in science organizations \citep{Merton68}, a ``cumulative advantage'' in bibliometrics
\citep{Simon1955,Price1976}, and as a ``capital gain'' in economics
\citep{Yakovenko2009}.  Preferential attachment in random networks naturally
explains power-law distributions \citep{BA99} of node sizes defined by the
number of links.  This approach explained power laws emerging across
different fields of science, e.g.  in the World Wide Web structure
\citep{AJB99}, protein interactions \citep{JMBO01}, metabolics
\citep{Ravasz+02}, transportation, and social networks, and scientific
collaborations \citep{Barabasi+02,NewmanCoauthorship2004}.  Here we describe a system of dense cores
growing in a molecular cloud by preferential attachment.  Gravitational
forces representing ``links'' between dense cores are distance-dependent,
hence we exploit the spatial network formalism \citep{Barthelemy11}.

When a new parcel emerges in the system, it becomes subject to multiple
competing attractive gravitational forces from existing dense core
progenitors, and at the same time to drag forces as it moves through the
ISM.  We assume that drag forces dominate over the inertia, so that the
exact dense core which will acquire a given parcel is determined only by the
competition of forces at the parcel's starting position (see
Fig.~\ref{fig_model}, right panel).  We set the parcel accretion probability
by a given core proportional to the initial gravitation acceleration toward
it (Fig.~\ref{fig_cartoon}, left panel).  In close vicinities of dense
cores, where the gravitational field is totally dominated by one mass, our
description becomes equivalent to the deterministic accretion onto that
particular core.  However, at the border separating areas of dominant
attraction (Fig.~\ref{fig_model}) from two cores, a parcel can be tipped
over it by stochastic pushes from other particles of the ISM.  The
probabilistic approach allows us to model that situation.  Hence, the
probability of a newly emergent parcel $j$ to join an existing dense core
$i$ is:

\begin{equation}
\pi_{j\to i}= \frac{1}{c_j} \frac{M_i}{|r_i-r_j|^\delta}
\end{equation}

\noindent where $M_i$ is the mass of the $i$th core, $|r_i-r_j|$ is the Euclidean
distance between the points, $\delta$ is the gravitational law exponent
($\delta=2$ in the 3-dimensional space). 
A possible alternative choice is $\delta=1$ which corresponds to the decay
of gravitational \emph{potential} rather than acceleration.  This, however,
is not very important for the preferential attachment description. We choose a
normalization so that the probability of a parcel joining \textit{some}
dense core is unity by using the continuous random fractal approximation to
sum over all cores:

\begin{flalign}
c_j=\sum\limits_{i} \frac{M_i}{|r_i-r_j|^\delta}=\int\limits_{l_i}^{L} dM(r) r^{-\delta} \nonumber \\
=\frac{M_\text{tot} D_m}{L^{D_m}} \frac{L^{D_m-\delta}-l_j^{D_m-\delta}}{D_m-\delta}
\label{eqn_norm}
\end{flalign}

\noindent where $l_j$ is a distance from a new parcel to the nearest dense
core.  Assuming $D_m>\delta$ and $l_j\ll L$, we can neglect $l_j$ so that
the statistics is dominated by distant dense cores, owing to the long distance
nature of gravitation.  Thus, the normalization factor is the same for all
parcels, regardless of where they emerge.

From examining the integrals in Eqs.~\ref{eqn_norm} and \ref{growtheqn} below
we can see that distant spherical layers of the ISM contribute very little to
the dense core growth or the parcel accretion, since their contribution is
proportional to $r^{D_m-1}/r^\delta$ or $r^{D_p-1}/r^\delta$ and vanishes in
the limit of large $r$.

The average dense core mass increase per time step is a probability weighted
sum over all possible positions where parcels emerge.

\begin{align}
&dM_i=dm\sum\limits_{\vec{r}} N(\vec{r})\pi_{j\to i} \nonumber \\
&=dm\int\limits_{l_i}^{L} D_p dr\frac{r^{D_p-1}}{L^{D_p}} \frac{L^\delta}{M_\text{tot}} \frac{D_m-\delta}{D_m} \frac{M_i}{r^\delta}=\beta \frac{M_i}{M_\text{tot}} dm
\label{growtheqn}
\end{align}

We notice the difference between our calculated accretion rate for
an individual core $dM_i/dt \propto M_i$ and the classical
Bondi--Hoyle--Lyttleton model \citep{HL39,Bondi52} of spherical accretion
($dM/dt \propto M^2$).  This inconsistency is trivially explained by the two
facts: (a) we assume a fractal distribution of the infalling matter so that
a thin spherical layer \emph{no longer contains} the mass $\rho R^2 dR$
used in the Hoyle--Lyttleton calculations; and (b) our model considers motions
of fractally distributed gas parcels in space to be overdamped as opposed
to ballistic motions, therefore the whole orbital computation including the
impact parameter and the escape velocity from \citet{HL39} is not applicable
to our case.

We denote the growth exponent $\beta =(1-\delta/D_m)/(1-\delta/D_p)<1$.  It
characterizes the growth rate of individual masses.  We can illustrate its
contribution to the dense core growth by using a simple, but manifestly
unrealistic assumption of a constant accretion rate in the system.  If at some
moment $t_0$, a dense core has mass $M_{i,0}$ and the global mass growth rate is
quasi-constant ($dm\sim dt$ and $M_\text{tot}\sim t$), then it grows in time
according to a sublinear power law:

\begin{equation}
M_i=M_{i,0} \left( \frac{t}{t_0} \right)^\beta
\end{equation}

In reality, this law does not hold because the global mass growth rate $K$
might not be constant.  However, one cannot directly observe the growth of a \emph{single}
dense core because it lasts tens of thousands of years.  As we show in the next
section, the directly observable quantity is a snapshot of the DCMF in the
$t \to \infty$ limit, which in turn is not affected by the specific time
dependence of $K$ and $F$.

\section{The power-law distribution from the master equation in networks}

The growth law for an individual dense core is not sufficient to derive the
mass distribution shape.  Therefore we use the master equation
\citep{DM02,newman2010networks} for the distribution evolution
\citep{Schnakenberg76} that describes probability flows between
different states of a system, in our case, different masses of dense cores
described by the DCMF $p(m,t)$.

The DCMF declines at low masses because dense cores cannot form below the
Jeans mass ($M_J$) where the gravitational contraction cannot overcome the
gas thermal pressure \citep{Jeans1901} or the turbulent support \citep{HC08}. 
In our model, dense core progenitors are generated across a finite range of
masses according to the initial probability distribution called source
function $f(m)$.

The three processes change the number of dense cores in a cell $m$ over one
time step $dt$: some cores of mass $m-dm$ grow and enter the cell, some
cores of mass $m$ grow and leave the cell, and $f(m)dt$ new cores are
created in this cell.  The accretion rate given by the growth equation
(Eq.~\ref{growtheqn}) is the same for all dense cores in a given cell. 
Putting these contributions together:

\begin{align}
&p(m,t+dt)(N+dN)=p(m,t)N \nonumber \\
&+\beta \frac{(m-dm)p(m-dm,t)N}{M_\text{tot}}-\beta \frac{mp(m,t)N}{M_\text{tot}}+f(m)dt
\label{eqn_master_discrete}
\end{align}

\noindent As the evolution runs for a long time, $p(m)$ converges to a
constant shape even if the number of dense cores and the total mass of the
system keep growing.  We take the dynamic equilibrium limit, so that
$p(m,t)\to p(m)$.  We also now go to the continuous-mass and continuous-time
description, such that $dm,dt\to 0$, while $dm/dt=K$.  In that case, we
replace the difference between the two accretion terms in
Eq.~\ref{eqn_master_discrete} above with a differential.

\begin{equation}
p(m)dN=-\beta dm \frac{N}{M_\text{tot}}\frac{d}{dm}(mp(m))+f(m)dt
\end{equation}

We can substitute $dN/dt=F$ and take the steady-state limit where $dm/dN
\to M_\text{tot}/N$.  This is actually a very weak assumption: we do not
presuppose any specific functional law for either $K=dm/dt$ or $F=dN/dt$, we
only need to assume that the ratio of those two rates is asymptotically
constant.  As both accretion \emph{and} generation of new dense core
progenitors are governed by the same physical processes, we expect them to
slow down at the same rate.  With this simplification, we obtain the growth
equation:

\begin{equation}
p(m)+\beta \frac{d}{dm}(mp(m))=\frac{f(m)}{F}
\label{masterseqn}
\end{equation}

\begin{figure}
\includegraphics[width=\hsize]{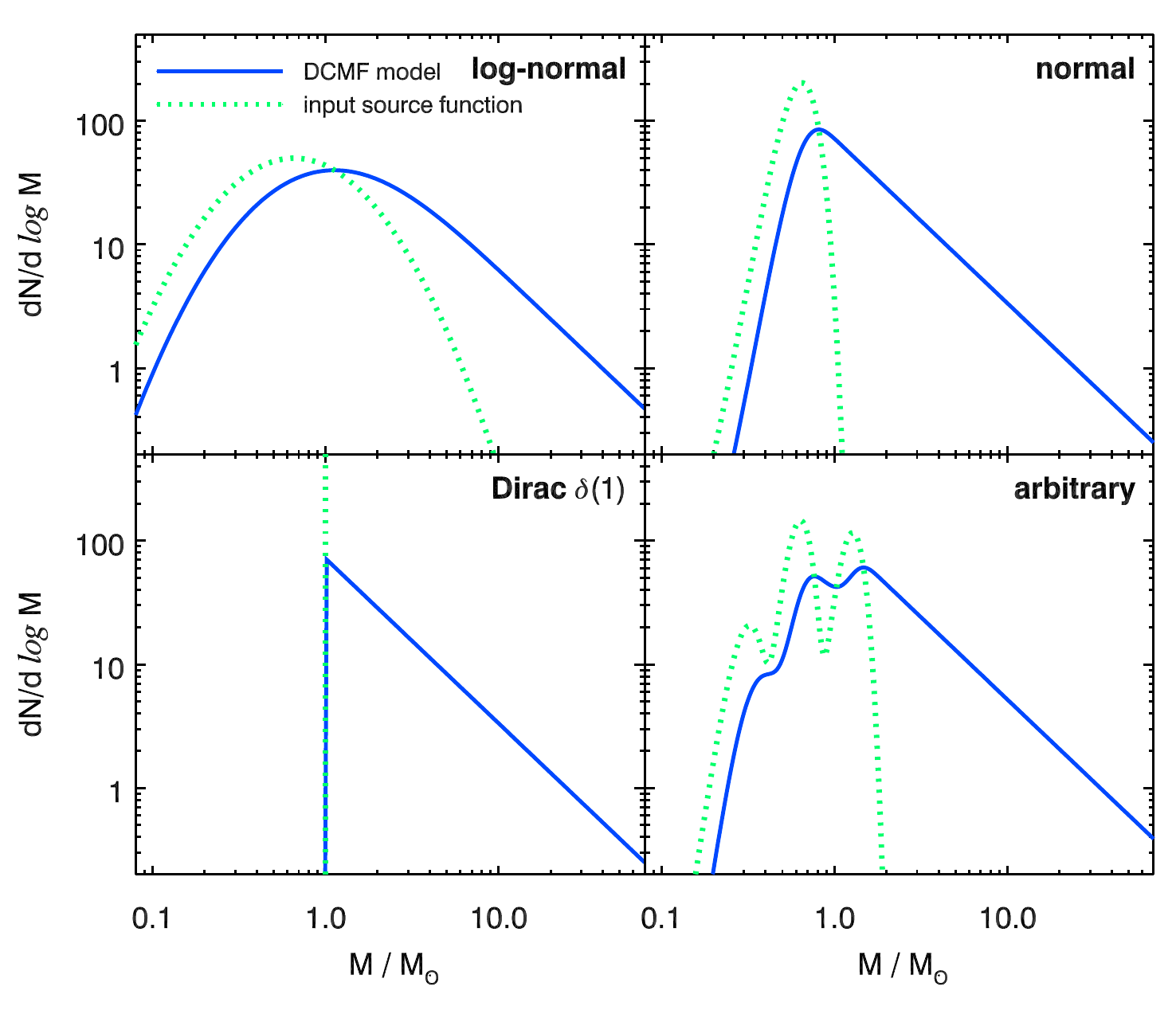}
\caption{Numerical solutions of Eq.~\ref{masterseqn} for different
shapes of the source function $f(m)$.  Top left: log-normal;
top right: normal (Gaussian); bottom left: Dirac $\delta$-function;
bottom right: an arbitrary multi-modal shape. We stress that for all
these source functions, the tail of $p(m)$ is a power law with the same
exponent as given by Eq.~\ref{eqnalpha}. \label{fig_source_function}}
\end{figure}

\begin{figure}
\includegraphics[width=\hsize]{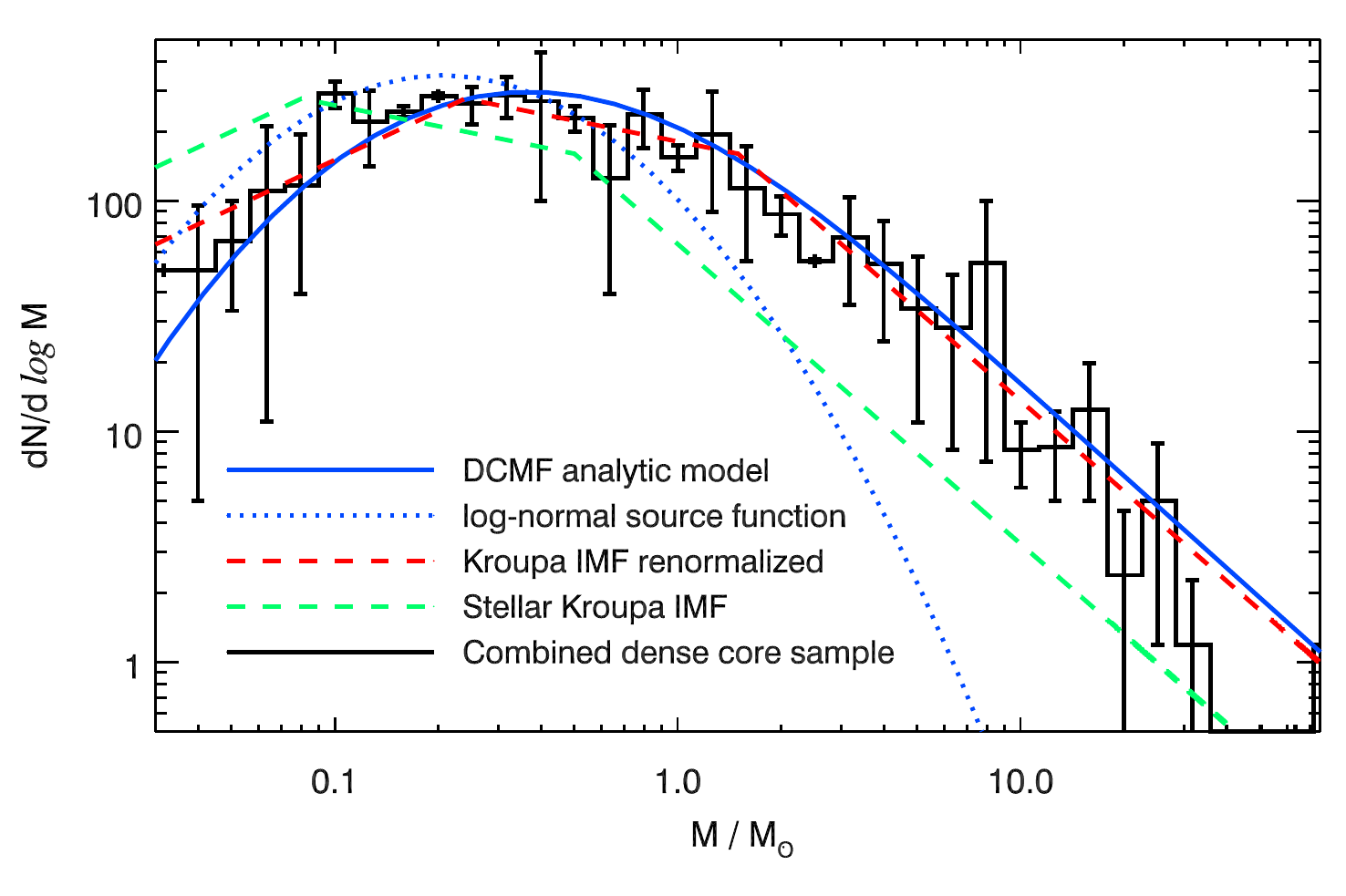}
\caption{An analytic DCMF model (blue solid line)
computed for fractal dimensionalities $D_m=2.35$, $D_p=2.5$, and a
log-normal source function (blue dotted line) are compared to
the Kroupa stellar IMF \citep{Kroupa01} (green dashed line), the Kroupa IMF shifted by a factor of
3 to higher masses (red dashed line), and an observed mass distribution of
dense cores in four star-forming regions (black histogram).  We co-added
observed mass distributions of 555 dense cores not containing protostars in the
Orion, Perseus, Ophiuchus, and Taurus star-forming regions \citep{Sadavoy+10}
by normalizing the numbers of cores in the $1.25<M<3.2 M_{\odot}$ mass
range. The uncertainties were estimated by varying dense core temperatures by
30\% \citep{Sadavoy+10}.
\label{fig_imf}}
\end{figure}

The Eq.~\ref{masterseqn} acts as a filter (a linear functional map)
that converts an initial density fluctuation spectrum $f(m)$ for dense core
progenitors into a DCMF $p(m)$.  Note that all time-dependent
quantities, such as $K$ or independently standing $F$ (now it only
normalizes the differential source function) have canceled out, thus the DCMF
shape does not depend on how the system slows down in time.
This equation preserves the normalization because
$\int\limits_{0}^{\infty}p(m)dm=\int\limits_{0}^{\infty}f(m)dm/F=1$.  An
exact analytical solution is only possible for some simple functional shape
of $f(m)$, but a number of numerical solutions are presented for
illustrative purposes in Fig.~\ref{fig_source_function}.  At high masses,
for any choice of $f(m)$ the DCMF develops the same power-law tail with an
exponent defined only by the fractal mass distribution properties, while at
low masses it essentially preserves the source function shape, with a smooth
transition in between.  In order to match observations, we take a
log-normal source function of a form $f(m)\propto \frac{1}{m}\exp \left(
-\frac{(\ln m - \mu)}{2 \sigma^2} \right)$ \citep{Chabrier03,HC08,HF12}. Its
maximum lies at $M_\text{max}=O(\exp (\mu))=O(M_J)$. 
In Fig.~\ref{fig_imf} we pickthe  $\mu$ and $\sigma$ that best resemble 
the observed distribution.

While the equation \ref{masterseqn} allows us to accurately match the
observed DCMF, its relevance and generality stretches beyond that.  To
calculate the high-mass tail of the distribution analytically, in that limit
we can neglect a rapidly decaying $f(m)$ (e.g.  a decaying exponent,
Gaussian, or log-normal).  Then, Eq.~\ref{masterseqn} becomes
homogeneous and has an analytic solution of a form $p(m)=Cm^{\alpha}$,
\emph{regardless of the $f(m)$ input source function shape}:

\begin{equation}
p(m)=Cm^{\alpha}, \, \alpha=-\left(1+\frac{1}{\beta} \right)=-\left(1+\frac{1-\delta/D_p}{1-\delta/D_m} \right)
\label{eqnalpha}
\end{equation}
\\
\section{Nonlinear accretion}

\cite{Maschberger13} considers the dense core growth through accretion
that is in general both nonlinear and stochastic.  The accretion rate
derived above is linear ($dM_i/dt\propto M_i$), although in general,
nonlinear cases are also possible with $dM_i/dt\propto M_i^a$ and $a\neq 1$. 
Accretion can be either sublinear ($a<1$, e.g.  \citealp{BCBP01}) or
superlinear ($a>1$, e.g.  $a=2$ in \citealp{Bondi52}).

\cite{Maschberger13} describes the process of competitive accretion
using the Stratonovich stochastic calculus formalism in order to predict the
mass distribution of dense cores when the accretion rate is partially random
and fluctuating.  The possible fluctuations need to be restricted to be 
exclusively
non-negative to rule out the mass loss.  In our theory, we account for the
stochasticity of accretion using the master equation (Eq.~\ref{masterseqn}). 
Since in our case, the only possible transition from each bin in mass is to the next bin,
our calculation is also restricted to non-negative fluctuations. Therefore, 
by plugging an alternative accretion rate $dM_i/dt\propto M_i^a$, the tail part of
the DCMF becomes:

\begin{equation}
p(m)=Cm^{-a}\exp \left( -\frac{1}{\beta} \frac{m^{1-a}}{1-a} \right)
\end{equation}

Here the constant $\beta$ is now dimensionful for $a\neq 1$. For sublinear 
accretion, the DCMF decays at high mass as stretched exponential, while for
the superlinear growth it results in a shallow power law $\simeq m^{-a}$.

Here the nonlinear accretion is directly analogous to the nonlinear
preferential attachment in network science
\citep{KRL00,newman2010networks}.  The sublinear preferential
attachment similarly results in a stretched exponential type distribution. 
The superlinear preferential attachment results in a situation where a few
network nodes accumulate a macroscopic fraction of all edges in the network. 
This issue is recognized in \cite{Maschberger13} as an ``explosion'' of
dense core masses in the absence of noise.  Since there is no observational
evidence of star-forming regions, where the entire mass is dominated by a
few very massive stars, the explosive growth scenario seems unrealistic.

Because the phenomenology of accretion in the fractal media is not clear,
we restrict the further analysis to the simple case of linear accretion
following from our model and given by Eq.~\ref{masterseqn}.

\section{Discussion and Summary}

\subsection{Dependence of the slope $\alpha$ on input parameters}

Having obtained an analytic expression (Eq.~\ref{eqnalpha}) for the
power-law exponent $\alpha$, we can now explore how it behaves as we vary the
three possible parameters of the system $D_m$, $D_p$ and $\delta$.  An
important thing to stress is that none of the three parameters bears any
dimensional units.  On one hand, this is due to the term ``scale-free
distribution'': in the high-mass tail $p(m)\sim Cm^\alpha$ there is no
characteristic mass or scale that defines the shape of the distribution (as
opposed to other functional forms, such as normal, log-normal, or
exponential).  On the other hand, these parameters are directly related to
fundamental scaling laws of statistical physics relevant on a broad spectrum
of length, time, and mass scales.

By substituting the observed value $D_m=2.35$ \citep{FPW91,EF96}, we obtain $\alpha = -(1
+ 6.7 (1-2/D_p))$.  The uniform density distribution of gas parcels
corresponds to $D_p=3$ and $\alpha \approx -3.24$.  The Salpeter value
$\alpha=-2.35$ corresponds to $D_p=2.5$.  This fractal dimensionality is
predicted and observed in a number of physical systems governed by Brownian
processes such as the diffusion-limited aggregation \citep{Meakin83} known
to take place for dust in the ISM \citep{2011A&A...536A..24P}. Whether or
not an actual Brownian process stays behind the $D_p$ value is
beyond the scope of our work, however, finding how observed physical 
properties of the ISM may affect its fractal dimensionality at the low-mass 
end might provide a clue to our understanding of IMF variations.

We also notice that if $D_p=D_m$, the second term in the expression turns
into unity and yields $\alpha=-2.0$.  For generic values of the parameters,
the numerator in the last fraction of Eq.~\ref{eqnalpha} represents the
variety of choice of parcels for a given dense core (Fig.~\ref{fig_cartoon}
(right)), whereas the denominator represents the variety of choices of dense
cores for a parcel (Fig.~\ref{fig_cartoon} (left)).  If both are distributed
in space following the same fractal dimension, then the patterns of parcels
attaching to dense cores no longer depend on the spatial coordinates. 
Effectively, for $D_p=D_m$ the spatiality of the problem cancels out and
it reduces to the ``regular'' preferential attachment process (avoiding the
double-counting of network edges as in \citealp{BA99}).  The expression also
becomes independent of $\delta$, thus removing the necessity of our model
choice to weigh the accretion probabilities by gravitational accelerations
or gravitational potentials.

A specific possible value of $\alpha=-2.0$ at $D_p=D_m$ is described by
\cite{Hopkins2013} as ``a generic scale-free distribution, allotting equal
mass to each equal logarithmic interval in mass.''  The actual value
obtained by Hopkins is equal to $2$ plus a small addition coming from
various effects related to the properties of the turbulent ISM and magnetic
fields.  In our theory, that addition appears naturally from considering a
two-phase medium, i.e.  different spatial distributions of dense cores and
parcels.

Our result favors the unimodal IMF shape over the bimodal.  The broken power 
law
\citep{Kroupa01} is acceptable as a fitting approximation for the smooth
transition between the low-mass decline and the high-mass power-law tail. 
This agrees with the conclusions drawn from numerical simulations
\citep{Elmegreen97,Elmegreen02} of the fractal ISM evolution. 
\citet{Clauset2009} specifically discuss the difficulties of fitting
power laws and other fat-tail distributions to empirical data and point out
that it is often hard to distinguish which model represents the data better.

Then, given no evidence that the turbulence induced ISM fractal
dimensionality $D_m = 2.35$ varies across different star-forming regions,
the variation of $2 < D_p \le 3$ remains the only channel to explain
possible IMF non-universality. The two hard limits are $\alpha=1$ ($D_p=2$) and
$\alpha=3.24$ ($D_p=3$).

For a system with a finite number of cores $N$ we estimate the mass ratio of
the largest to smallest cores in the power-law regime as
$M_\text{max}=M_\text{min} N^\beta$.  This explains the observed correlation
between the most massive star mass and the total star cluster mass that we
can calculate for any specific solution of Eq.~\ref{masterseqn}
\citep{Kroupa+13}.

\citet{Larson92} attempted to relate the ISM fractal dimensionality to
the IMF shape by assuming that the entire mass from some fragment of the
molecular cloud surrounding a core accretes onto it.  Then, the IMF power-law
exponent becomes equal to the fractal dimensionality.  In our model,
however, we do not make the assumption that every core grows by accreting
matter from a distinct region of the cloud but rather consider the
competitive accretion (or preferential attachment) in order to account for
overlapping basins of gravitational attraction.

A filamentary distribution of parcels will correspond to $1<D_p<2$. This
will change the convergence of integrals in Eq.~\ref{eqn_norm}--\ref{growtheqn} 
but will still result in a power-law mass function.  In principle, it is
possible to introduce a scale-dependent fractal structure where $D_p$ and/or
$D_m$ change at some characteristic scale $R_f$.  This will, however, make
the calculations bulky and will also introduce additional free parameters so
that the solution behavior will be more difficult to investigate and
explain analytically.

\subsection{Bottom-heavy mass functions}
\begin{figure}
\includegraphics[width=\hsize]{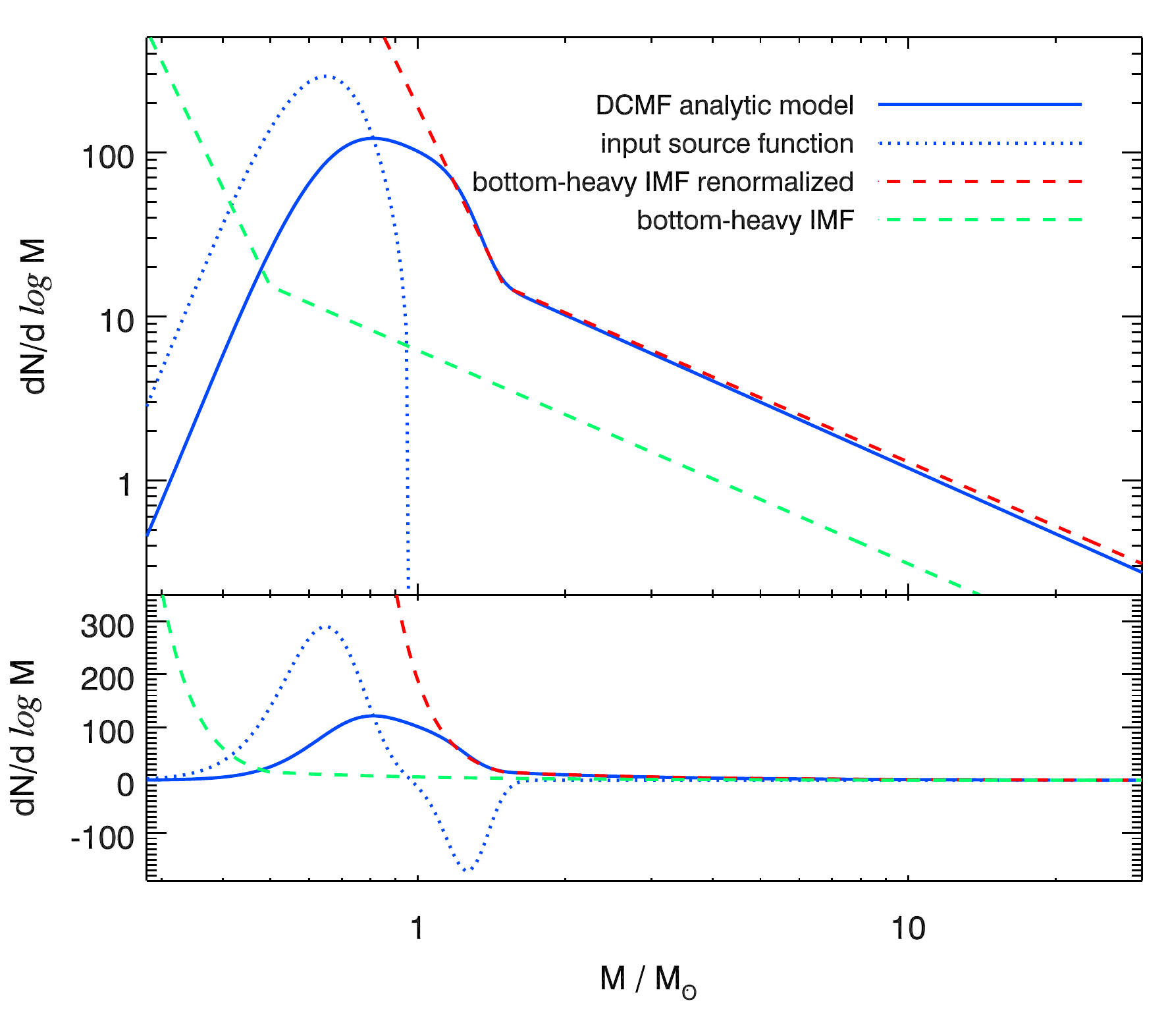}
\caption{{\bf (Top)} An analytic DCMF model (blue solid line)
computed for fractal dimensionalities $D_m=2.35$, $D_p=2.5$, and a
source function having negative values at $M>0.97 M_{\odot}$ (blue dotted 
line) are compared to a fiducial bottom-heavy stellar IMF (green dashed line),
and the same IMF shifted by a factor of 3 to higher masses (red dashed
line). {\bf (Bottom)} The vertical axis scale is linear 
in order to demonstrate the partially negative source function shape.
\label{figbottomheavy}}
\end{figure}

The low-mass star dominated bottom-heavy IMF suggested by
recent observations \citep{vDC10,Cappellari+12} has a
slope at certain masses steeper than the asymptotic value $\alpha$ 
(Fig.~\ref{figbottomheavy}).  The logarithmic slope is given by 
$d\log p(m)/d\log m=(dp(m)/dm)(m/p(m))$. We derive it directly from the 
master equation
(Eq.~\ref{masterseqn}) in a self-referential form, without solving it for any
specific $f(m)$:

\begin{equation}
\frac{d\log p(m)}{d\log m}=\frac{dp(m)}{dm}\frac{m}{p(m)}=\alpha + 
\frac{f(m)}{\beta F p(m)}
\end{equation}

\noindent $f(m)$ is always non-negative because dense cores in our model are
never destroyed, and $p(m)$ is non-negative as a
probability distribution.  Asymptotically, their ratio $f(m)/p(m)\to 0$ 
because $f(m)$ has exponential or faster decay and $p(m)$ is a ``slower''
power law.  Thus, the logarithmic slope $d\log p(m)/d\log m \to
\alpha$ from above, and it can never become steeper than $\alpha$ unless 
the
non-negativity condition is violated.  We solved Eq.~\ref{masterseqn}
for a fiducial source function that is negative 
for some masses (Fig.~\ref{figbottomheavy}) in order to illustrate how a
bottom-heavy DCMF can be established. Because $\alpha$ only depends 
on the fundamental scaling exponents $\delta$, $D_m$ and $D_p$, and because
it serves as a hard lower bound on the DCMF slope, bottom-heavy mass
functions are ruled out by our theory for the linear accretion regime.

This conclusion comes into tension with the results that suggest
a bottom-heavy IMF shape in elliptical galaxies \citep{vDC10,Cappellari+12}
with $\alpha \approx 3$.  It is worth mentioning, that those conclusions have 
recently been challenged by statistical data analysis
\citep{Smith14,CSF15}, observations of extragalactic X-ray binaries
\citep{Peacock+14}, and strong gravitational lensing \citep{SLC15}. 
However, one has to keep in mind that it is impossible to
observationally measure the IMF slope at masses ($M>0.7 M_{\odot}$) in old
stellar populations, because stars at that mass range have already evolved
into remnants.  Therefore, the unimodal steep IMF slope cannot be excluded
as a solution satisfying both our theory and observations, if future studies
explain how the parcel fractal density dimensionality $D_p$ depends on a
galaxy mass or the ISM metal content.

Also, if we admit variations of the unimodal IMF slope across different 
star-forming regions in the same galaxy, the observational appearance of an
IMF to be bottom-heavy becomes plausible for composite stellar populations
(e.g.  galaxies formed by major dry mergers).  That can happen if, for
example, a combined IMF shape is determined for a stellar system that
consists of several building blocks having different intrinsic IMF slopes
and comparable masses. Then, the combined stellar distribution will be
dominated by low-mass stars preferably from a bottom-heavy building block,
while its high-mass end will be defined by a shallow (top-heavy) IMF stellar
component.  This explains why until now, no stand-alone star cluster or a
star-forming complex with a bottom-heavy IMF has been found 
with the same integrated light spectral diagnostics as those used to derive the
bottom-heavy IMF shape in giant early-type galaxies \citep{vDC10,vDC11}. 
Stellar systems that can be reasonably well represented by simple stellar
populations, such as ultracompact dwarf galaxies and massive globular
clusters, exhibit stellar masses corresponding to the low-mass IMF slopes
between Kroupa and Salpeter \citep{CMHI11,vDC11,PCK13}.

\subsection{Summary}

We presented a simple analytic approach that addresses the following major
points formulated in the \emph{introduction} and left unexplained by
existing IMF theories: 
\begin{itemize}
\item[i] The scale-free distribution of dense cores by
mass is established by the process of preferential attachment (competitive
accretion) of mass onto dense cores. 
\item[ii] When the system mass grows, the
distribution shape asymptotically stabilizes. 
\item[iii]  The power-law exponent
depends only on two parameters, $D_m$ and $D_p$, fractal dimensions of the
turbulent ISM and accreting gas, directly connected to their fundamental
physical properties.
\end{itemize}

Our theory relies on the qualitative 
description of the supersonic turbulence that follows from the basic
Kolmogorov theory. The real structure of the supersonic turbulent flow 
in the ISM might be different and will potentially affect our results.
However, if the density distribution can still be described as fractal, it
will only affect the power-law slope as suggested by Eq.~\ref{eqnalpha}.
We explain the bimodality of the Kroupa IMF as a result of a two-component
fitting of the intrinsically unimodal distribution in the transition region
($0.08 < M < 0.5 M_{\odot}$) between a power law at high masses and a
declining part at low masses (Fig.~\ref{fig_imf}).  By our calculation of the lower
bound on the logarithmic slope, in the transition region it should never be
steeper than that at higher masses, therefore we rule out a bottom-heavy IMF
shape for any single stand-alone star-forming region. 

\begin{acknowledgments}
We are grateful to our anonymous reviewers whose feedback helped
us to improve the manuscript.
IC acknowledges the support from the Telescope Data Center, Smithsonian
Astrophysical Observatory.  His theoretical and observational IMF studies are
supported by the Russian Foundation for Basic Research projects 15-52-15050
and 15-32-21062, and the President of Russian Federation grant
MD-7355.2015.2.  The idea of applying network science formalism to the IMF
theory was developed by the authors during the annual Chamonix workshop in
2014 supported by the Russian Science Foundation project 14-22-00041.  The 
authors
appreciate fruitful discussions with and useful suggestions from J.~Silk,
P.~Hennebelle, G.~Mamon, P.~Kroupa, S.~Mieske, M.~Kurtz, I.~Zolotukhin,
C.~Lada, and M.~E.~J.~Newman.  
\end{acknowledgments}

\bibliographystyle{apj}
\bibliography{DCMF}

\end{document}